\documentclass[onecolumn,pra,aps,amsmath,amssymb]{revtex4}
\usepackage{units}
\usepackage{color}
\usepackage{dsfont}

\newcommand{\be}{\begin{eqnarray}}
\newcommand{\ee}{\end{eqnarray}}

\begin{document}

%\author{Harald Wunderlich$^{a,b}$$^{\ast}$\thanks{$^\ast$Corresponding author. Email: h.wunderlich@physik.uni-siegen.de
%\vspace{6pt}} and Martin B. Plenio$^{b,c}$\\\vspace{6pt}  $^{a}${\em{
%Fachbereich Physik, Universit\"at Siegen, 57068 Siegen, Germany}};
%$^{b}${\em{Institute for Mathematical Sciences, Imperial College London,
%53 Prince's Gate, SW7 2PG London, UK}}; 
%$^{c}${\em{QOLS, Blackett Laboratory, Imperial College London, Prince Consort
%Road, SW7 2BW London, UK}};\\\vspace{6pt}\received{released February 2009} }

\author{Harald Wunderlich$^{1,2}$}

 \email{h.wunderlich@physik.uni-siegen.de}

\author{Martin B. Plenio$^{2,3}$}

\affiliation{$^{1}$Fachbereich Physik, Universit\"at Siegen, 57068 Siegen,
Germany \\
$^{2}$Institute for Mathematical Sciences, Imperial College London,
53 Prince's Gate, SW7 2PG London, UK
\\
$^{3}$QOLS, Blackett Laboratory, Imperial College London, Prince Consort
Road, SW7 2BW London, UK}
 
%\author{Martin B. Plenio}
% \email{m.plenio@imperial.ac.uk}
%\affiliation{Institute for Mathematical Sciences, Imperial College London,
%53 Prince's Gate, SW7 2PG London, UK}
%\affiliation{QOLS, Blackett Laboratory, Imperial College London, Prince Consort
%Road, SW7 2BW London, UK}

\title{Quantitative verification of entanglement and fidelities
from incomplete measurement
data}

\begin{abstract}
Many experiments in quantum information aim at creating multi-partite
entangled states. Quantifying the amount of entanglement that was
actually generated can, in principle, be accomplished using 
full-state tomography. This method requires the determination 
of a parameter set that is growing exponentially with the number 
of qubits and becomes infeasible even for moderate numbers of
particles. Non-trivial bounds on 
experimentally prepared entanglement can however be obtained 
from partial information on the density matrix.
 As introduced in [K.M.R. Audenaert and M.B. Plenio, New J. Phys. 
 \textbf{8}, 266 (2006)], the fundamental question is 
then formulated as: What is the entanglement content of the 
least entangled quantum state that is compatible with the 
available measurement data? 

We formulate the problem mathematically \cite{Audenaert06}
employing methods from the theory of semi-definite programming 
and then address 
this problem for the case, where the goal of the experiment 
is the creation of graph states. The observables that we consider 
are the generators of the stabilizer group, thus the number of 
measurement settings grows only linearly in the number of 
qubits. We provide analytical solutions as well as numerical 
methods that may be applied directly to experiments, and 
compare the obtained bounds with results from full-state 
tomography for simulated data. 
\end{abstract}

%\begin{keywords}
%entanglement estimation, fidelity, robustness, stabilizer states
%\end{keywords}

\maketitle

\section{Introduction}
Detecting and quantifying entanglement is one of the core problems 
in quantum information theory \cite{Virmani07, Horodecki07}. The 
detection of entanglement can in principle be accomplished by 
measuring the complete quantum state and, thereafter, applying 
separability tests. However, the dimension of the density matrix 
grows exponentially with the number of constituents of the system. 
Therefore, full state tomography becomes very costly and 
experimentally infeasible already for a moderate number of 
particles. Thus, it is of interest to detect and quantify 
entanglement, even when only partial information on the density 
matrix is known. Entanglement Witnesses represent one way to verify 
the existence of entanglement with only a few measurements
\cite{HorodeckiHH96}. In particular, witnesses for the detection of stabilizer states have been constructed in \cite{Toth05}, which also provide lower bounds on the fidelity.

In \cite{Brandao05} it was demonstrated that one may
define an entanglement measure on the basis of witness operators
and provide lower bounds on entanglement measures 
\cite{Eisert07-9,Guehne07}. 

However, the restriction to witness operators is unnecessary
and may neglect information that is obtained when measuring
the local operators into which the witness operator has been 
decomposed \cite{Audenaert06}. A direct calculation of 
the least amount of entanglement (in accordance with an 
entanglement measure of choice) that is compatible with the 
measured data of {\em arbitrary} observables is proposed in 
\cite{Audenaert06} and this approach is guaranteed to deliver 
the best lower bounds that can be obtained from the 
information that is available. The same philosophy may also be 
followed when bounding other quantities such as the fidelity with 
a desired target state. We apply the method described in 
\cite{Audenaert06} to the case where the goal of the experiment 
was the creation of cluster states. The observables we consider 
are the generators of the stabilizer group (for an introduction 
to the stabilizer formalism see e.g. \cite{Gottesman97}). Thus 
the number of measurement settings grows only linearly in the 
number of qubits.

This article is structured as follows: First, we describe how to
provide lower bounds on the fidelity with a target state (here 
stabilizer states) in Sec. \ref{sec:MinimalFidelity} and in 
Sec. \ref{sec:ExamplesMinimalFidelity} apply this to some simple 
examples. Sec. \ref{sec:Entanglement-Measures} discusses the general 
approach to estimate robustness measures from incomplete information 
on the density matrix. Then, in Sec. \ref{sec:Application} we 
utilize this approach to obtain lower bounds on the Global 
Robustness of Entanglement, and give closed formulae for systems
consisting of two, three, and four qubits. A comparison of the 
obtained bounds with exact values for noisy cluster states is 
provided in Sec. \ref{sec:Quality}.

\section{Minimal Fidelity and Entanglement}
\label{sec:MinimalFidelity}
In many experiments we aim at creating a particular
pure quantum state $|\phi\rangle$. Needless to say, experimental
imperfections and noise will usually lead to a noisy approximation
to this state, i.e. a fidelity that is different
from unity. This naturally raises the question as to how
close we actually are to the target state. 
It is desirable to find simple
sets of measurements that give us enough information
to find useful lower bounds on the fidelity that has been
achieved in the experiment. This problem may be solved
with the methods that have been developed earlier in \cite{Audenaert06}.
More formally, we will measure a set of observables $\{A_i\}$
and find measured mean values $a_i$. Then we will find the
state $\rho$ that predicts the mean values $a_i$ and that has the
least fidelity with the target state $|\phi\rangle$. Mathematically
this is formulated as
\be
F_{min} &  = & min[tr[|\phi\rangle\langle\phi| \rho] : tr[A_i \rho] = a_i, \rho \ge 0] .
\ee
The solution to this problem is called the primal optimal.
This problem is in fact numerically very efficiently solvable
as it is linear program which is a special case of a
semi-definite program. As such there are firstly very efficient
numerical algorithms and, employing the concept
of duality, one can also find lower bounds on the minimization
problem \cite{Boyd04}. Indeed, by duality we find that
\be
F_{min} & = & 
%max [min[tr[(|\phi\rangle\langle\phi| -\sum_i \lambda_i A_i)\rho]
%+\lambda_i a_i]\\
max [\sum_i \lambda_i a_i: (|\phi \rangle \langle \phi|-\sum_i \lambda_i
A_i) \ge 0].
\ee
The solution to the latter, dual, problem is the 
dual optimal. That we really have equality, as we have implied
here, is not trivial, but is true for linear programs, and for 
general semi-definite programs it is true under very mild 
conditions (see \cite{Boyd04} for details) and is usually 
safe to assume at the beginning (though that this needs to be 
checked should be remembered when primal and dual optimum do
not appear to coincide).
The same methods can also be used to verify and quantify
entanglement measures. In the bi-partite setting this
can for example be done for the logarithmic negativity 
\cite{Virmani07,Werner02-65, Plenio05-95,Eisert02} and simple 
analytical formulae can be given \cite{Audenaert06} for 
useful sets of observables. For multi-partite
settings we have a variety of measures available \cite{Virmani07, Horodecki07}. One
may chose simple generalizations of the negativity measures
but may also study robustness measures \cite{Vidal99}. The method for the description
of negativity measures can be deduced directly from \cite{Audenaert06}. In this note we will
thus only present the basic approach for robustness measures.
\section{Examples: Fidelity Estimation}
\label{sec:ExamplesMinimalFidelity}
\subsection{GHZ States}
The optimization problem formulated above may look somewhat 
daunting. Let us therefore consider some examples. First, we 
consider the quantitative verification of the fidelity with the EPR state 
$|\phi_2\rangle = \frac{1}{\sqrt{2}}(|00\rangle + |11\rangle)$. 
Let us measure the expectation values of the observables 
$A_1 = X_1 \otimes X_2$ and $A_2 = Z_1 \otimes Z_2$ where $X$ 
($Z$) is the Pauli $x$ ($z$) operator. The unit trace condition 
on the density matrix is $tr[\rho]$ so that we have 
$A_3 = \mathds{1}$. Then we find
\be
F_{min} = \frac{a_1 + a_2}{2}.
\ee
This is seen from the choice
\be
\rho = \frac{1}{4}
\begin{pmatrix}
1 + a_2 & 0 & 0 & 2a_1 + a_2 - 1
\\
0 & 1 - a_2 & 1 - a_2 & 0
\\
0 & 1 - a_2 & 1 - a_2 & 0
\\
2a_1 + a_2 - 1 & 0 & 0 & 1 + a_2
\end{pmatrix}
\ee
and for the dual problem with the choice $\lambda_1 = \lambda_2 = \frac{1}{2}$
and $\lambda_3 = 0$.
The verification of the GHZ-fidelity, that is the overlap
with the state $|\phi\rangle = \frac{1}{2}(|000\rangle + |111\rangle)$ may also be
considered. Here we measure the observables
$A_1 =  X_1 \otimes X_2 \otimes X_3$, $A_2 = Z_1 \otimes Z_2 \otimes \mathds{1}$, and $A_3 = \mathds{1} \otimes Z_2 \otimes Z_3$.
Again, $A_4 = \mathds{1}$. Here it is a little harder to find the closed
formula but it is actually
\be
F_{min} = max[\frac{a_1 + a_2 + a_3 - 1}{2}, 0].
\ee
The dual optimal is then of course $\lambda_1 = \lambda_2 = \lambda_3 = \frac{1}{2}$
and $\lambda_4 = -\frac{1}{2}$. The optimal $\rho$ for the primal problem
has $\rho_{1,1} = max[(a_2 + a_3)/4, 0]$, $\rho_{i,j} = \rho_{i,i} = \rho_{j,j} = \rho_{j,i}$ for $i, j \in [2, . . . , 7]$ and $\rho_{1,8}$ such that the above optimal emerges. 

For a general n-particle GHZ-state we measure for example $A_1 = X_1 \otimes \dots \otimes X_n$, and $A_m = Z_{m-1} Z_m$ for $m\ge 2$, to find
\be
F_{min} = max[\frac{a_1 + . . . + a_n - n + 2}{2}, 0]
\ee
It is straightforward to read off the form of the dual optimal
from this expression.
\subsection{Cluster States}
We find the same bounds on the fidelity for cluster states. As a matter of
fact, these fidelity estimates are true for any observables which generate
a stabilizer group. A proof is given in Appendix 
\ref{sec:proof-for-fidelity}.

\section{Entanglement Measures}
\label{sec:Entanglement-Measures}
In this section we will discuss the estimation of entanglement
measures from tomographically incomplete measurements . As mentioned, \cite{Audenaert06} discusses already the logarithmic
negativity measures. For example, when one measures $X \otimes X$ and $Z\otimes
Z$ and finds $a_x$ and $a_z$ then as demonstrated in \cite{Audenaert06}: $E_{min} = max(0,log(|a_x|+|a_z|))$.
%\be
%E_{min} = max(0,log(|a_x|+|a_z|))
%\ee
and if one additionally measures $Y \otimes Y$ and finds $a_y$, then $E_{min} = max(0,log(1 + |a_x| + |a_y| + |a_z|))$.
%\be
%E_{min} = max(0,log(1 + |a_x| + |a_y| + |a_z|)).
%\ee
Here we will present the approach for the Global Robustness of Entanglement. For bi-partite systems, this is defined as
\be
E(\rho) = min[tr[\sigma] : \sigma \ge 0, \rho^\Gamma+\sigma^\Gamma \ge 0]
\ee
where $\sigma$ must be Hermitian and positive-semidefinite, and $\Gamma$
denotes partial transposition. For many particles, say $n$, a natural extension is
\be
E(\rho) = min[tr[\sigma] : \sigma \ge 0, \rho^{\Gamma_{\alpha}}+\sigma^{\Gamma_{\alpha}} \ge 0, 
\forall \alpha \in \{1,\dots,n\}].
\ee
Again, given some expectation values $tr[\rho A_i] = a_i$, we
would then determine
\be
E_{min} = min[E(\rho): tr(\rho A_i)=a_i, \rho \ge 0].
\ee
This is again a semi-definite program and is thus rapidly
solvable using numerical programs. For analytical work
it will again be interesting to derive the dual which will
allow us to find lower bounds on the above minimization.
This derivation  can be done in the following steps:
\be
\nonumber
E_{min} & = & min[tr[\sigma] : tr(\rho A_i)=a_i, 
\rho^{\Gamma_{\alpha}}+\sigma^{\Gamma_{\alpha}} \ge 0, \rho \ge 0,
\sigma \ge 0]
\\
\nonumber
%& = & max[min\{tr[\sigma]-\sum_{\alpha} tr[\eta_{\alpha} 
%(\rho^{\Gamma_{\alpha}}+\sigma^{\Gamma_{\alpha}})]\\
%\nonumber
%& & -\sum_i \mu_i (tr[\rho A_i]-a_i):\rho,\sigma \ge 0 \}: \eta_{\alpha} \ge
%0, \mu]
%\\
%\nonumber
& = & max[min\{tr[\sigma (\mathds{1}-\sum_{\alpha} \eta_{\alpha}^{\Gamma_{\alpha}})]-\sum_{\alpha}
tr[\eta_{\alpha}^{\Gamma_{\alpha}} \rho]
\\
\nonumber
& & -\sum_i \mu_i (tr[\rho A_i]-a_i):\rho,\sigma \ge 0 \}: \eta_{\alpha} \ge
0, \mu]
\\
\nonumber
& = & max[min\{-\sum_{\alpha} tr[\eta_{\alpha}^{\Gamma_{\alpha}} \rho] -\sum_i \mu_i (tr[\rho A_i]-a_i):\rho\ge
0\} :\eta_{\alpha} \ge 0, \mathds{1} \ge \sum_{\alpha} \eta_{\alpha}^{\Gamma_{\alpha}}, \mu]
\\
\label{eq:dualopt}
& = & max[\sum_i \mu_i a_i : \mathds{1} \ge \sum_{\alpha} \eta_{\alpha}^{\Gamma_{\alpha}},
\sum_{\alpha} \eta_{\alpha}^{\Gamma_{\alpha}}+\sum_i \mu_i A_i \le 0, 
\eta_{\alpha} \ge 0, \mu]
\ee

\section{Application to Stabilizer States}
\label{sec:Application}
In this section we will utilize the approach described in the previous section
to explicitly calculate lower bounds on the Global Robustness of Entanglement.
We will see that the proper choice of observables transforms the optimization
problem into a linear program, which may be solved analytically as well as
numerically using well-known algorithms like the Simplex method.

We assume now that the goal of the experiment was to create either a cluster state with the associated adjacency matrix $\Gamma_A$ or a GHZ state. Then a natural choice for the observables $A_i$ would be the generators $K_i$ of the abelian stabilizer group. For cluster states the stabilizers (or correlation
operators) are given by
%\begin{equation}
$K_i = X_i Z_{N_i}$
%\end{equation}
where the subscript $N_i$ is to be understood as applying $Z$-operators to
all neighbors of the $i$-th qubit in the lattice defined by $\Gamma_A$.
For $N$-qubit GHZ states the generators of the stabilizer group are:
%\begin{equation}
$K_1 = X_1 \otimes ... \otimes X_N$  
%\end{equation}
and for $k = 2, ... , N$:
%\begin{equation}
$K_k = Z_{k-1} Z_k$.
%\end{equation}
Due to the commutation relations fulfilled by these operators, it is easy to see that the symmetries that leave the observables $a_i = tr(\rho K_i)$
invariant are given by the transformation $\rho \longrightarrow \rho' = \frac{1}{2^N}\sum_{i_1,...,i_N=0}^1 K_1^{i_1}...K_N^{i_N}
\rho K_1^{i_1}...K_N^{i_N}$.
%with $i_1,...,i_N \in \{0, 1\}$. 
We may therefore
restrict our attention to states of the form:
\begin{equation}
\label{eq:rhosym}
\rho = \sum_{i_1,...,i_N = 0}^1 c_{i_1 ... i_N} K_1^{i_1} ... K_N^{i_N}
\end{equation}
with real coefficients $c_{i_1 ... i_N}$.
We can further restrict the matrices $\eta_\alpha$ of the dual problem to
have the same symmetries as the states $\rho$. The eigenvalues of $\rho$
are: $\lambda_{j_1 ... j_N}(\rho) = \sum_{i_1,...,i_N = 0}^1 (-1)^{i_1 j_1} ...(-1)^{i_N
j_N} c_{i_1 ... i_N}$,
%\begin{equation}
%\lambda_{j_1 ... j_N}(\rho) = \sum_{i_1,...,i_N = 0}^1 (-1)^{i_1 j_1} ...(-1)^{i_N
%j_N} c_{i_1 ... i_N}
%\end{equation}
where the $j_k \in [0,1]$ form a binary index for $\lambda$. The symmetries obeyed by $\rho$ also imply that the (unnormalized)
state $\sigma$ has the same symmetries as (\ref{eq:rhosym}). This can be seen as follows:
One may define the completely positive map $\Lambda(\psi) = \frac{1}{2^N}\sum_{i_1,...,i_N = 0}^1 K_1^{i_1} ... K_N^{i_N} \psi
K_1^{i_1} ... K_N^{i_N}$.
%\be
%\Lambda(\psi) = \frac{1}{2^N}\sum_{i_1,...,i_N = 0}^1 K_1^{i_1} ... K_N^{i_N} \psi
%K_1^{i_1} ... K_N^{i_N}
%\ee
Then, assume we found $E_{min}$ and the corresponding operator $\sigma$,
such that $(\rho+\sigma)^{\Gamma_{\alpha}} \ge 0 ~\forall \alpha \in \{1,\dots,N\}$.
Since $\Lambda((\rho+\sigma)^{\Gamma_\alpha})=\rho^{\Gamma_\alpha}+\Lambda(\sigma)^{\Gamma_\alpha}$, one concludes that $\sigma$ must be invariant under rotations of the stabilizer group. Thus: $\sigma = \sum_{i_1,...,i_N = 0}^1 d_{i_1 ... i_N} K_1^{i_1} ... K_N^{i_N}$, with eigenvalues $\lambda_{j_1 ... j_N}(\sigma) = \sum_{i_1,...,i_N = 0}^1 (-1)^{i_1 j_1} ...(-1)^{i_N
j_N} d_{i_1 ... i_N}$.
%Therefore:
%\be
%\sigma = \sum_{i_1,...,i_N = 0}^1 d_{i_1 ... i_N} K_1^{i_1} ... K_N^{i_N}
%\ee
%and the eigenvectors of $\sigma$ are given by
%\be
%\lambda_{j_1 ... j_N}(\sigma) = \sum_{i_1,...,i_N = 0}^1 (-1)^{i_1 j_1} ...(-1)^{i_N
%j_N} d_{i_1 ... i_N}
%\ee

\subsection{2 Qubits}
\subsubsection{2 Qubits: Upper Bound}
In this section we will evaluate the entanglement for a composite 
system of two qubits, supposedly prepared as a cluster state. The 
measurements performed on this system result in $a_1 = tr(\rho K_1)$ 
and $a_2 = tr(\rho K_2) $, with $K_1 = X \otimes Z$ and 
$K_2 = Z \otimes X$. W.l.o.g. we restrict to the case of positive 
$a_i$ but write the solutions of the more general case of
arbitrary $a_i$. The primal problem reads
% \begin{align}
% minimize & \quad tr(\sigma) 
% \\
% subject~to & \quad (\rho+\sigma)^{\Gamma_1} \ge 0, \quad \rho \ge 0, \quad \sigma \ge 0,\quad  tr(\rho K_i) = a_i
% \end{align}
\begin{align}
min [tr(\sigma): (\rho+\sigma)^{\Gamma_1} \ge 0, \quad \rho \ge 0, \quad \sigma \ge 0,
\quad  tr(\rho K_i) = a_i]
\end{align}

We denote the eigenvalues of $\rho$ resp. of its transpose by:
\begin{align}
\lambda_{j_1 j_2} (\rho) & =  \sum_{i_1,i_2 = 0}^1 (-1)^{i_1 j_1} (-1)^{i_2
j_2} c_{i_1 i_2}
%\\
%\lambda_{j_1 j_2} (\sigma) & =  \sum_{i_1,i_2 = 0}^1 (-1)^{i_1 j_1} (-1)^{i_2
% j_2} d_{i_1 i_2}
\\
\lambda_{j_1 j_2} (\rho^{\Gamma_1}) & =  \sum_{i_1,i_2 = 0}^1 (-1)^{i_1 j_1} (-1)^{i_2
j_2} (-1)^{i_1 i_2} c_{i_1 i_2}
%\\
%\lambda_{j_1 j_2} (\sigma^{\Gamma_1}) & =  \sum_{i_1,i_2 = 0}^1 (-1)^{i_1 j_1} %(-1)^{i_2 j_2} (-1)^{i_1 i_2} d_{i_1 i_2}
\end{align} 
Eigenvalues of $\sigma$ and $\sigma^{\Gamma}$ are of the same form, and we denote the corresponding coefficients with $d_{i_1 i_2}$.
The coefficients $c_{00} = 1/4$, $c_{10} = a_1/4$, $c_{01} = a_2/4$ are given by normalization and measurement constraints. In the case $a_1+a_2 \le 1$, one may set $c_{11} = 0$, which is the coefficient that changes sign under partial transposition. Thus, $\sigma = 0$ in this case. Otherwise, upper bounds on $tr(\sigma)=4d_{00}$ are obtained by the choice  $d_{00}=(a_1+a_2-1)/4$, $d_{i_1 i_2} = -d_{00}/3$ else, and $c_{11} = -d_{00}$.
%Upper bounds on $tr(\sigma)=4d_{00}$ can be readily obtained from:
%\begin{align}
%4d_{00} \ge 4(c_{10}+c_{01}-c_{00}) = a_1+a_2-1 
%\end{align}
%by
%\begin{align}
%\lambda_{11}(\rho^{\Gamma_1}+\sigma^{\Gamma_1})+\lambda_{11}(\rho)+\lambda_{00}(\sigma)
%\ge 0
%\end{align}
%and
%\begin{align}
%d_{00} \ge -c_{00}+c_{10}+c_{01}+c_{11} +d_{10}+d_{01}+d_{11} 
%\end{align}
%by
%\begin{align}
%\lambda_{11}(\rho^{\Gamma_1}+\sigma^{\Gamma_1}) \ge 0 
%\end{align}
%From these two inequalities one can now extract the minimality condition:
%\begin{align}
%c_{11} = -(d_{10}+d_{01}+d_{11})
%\end{align}
%A legitimate choice is $d_{10}=d_{01}=d_{11}=-c_{11}/3$. 
%Furthermore, one finds from $\rho \ge 0$:
%\begin{align}
%c_{11} \ge -c_{00}+|c_{10}|+|c_{01}|
%\end{align}
%There are two cases to be distinguished. If $a_1+a_2 \le 1$, then one may
%set $d_{i_1 i_2} = 0$, $\forall~i_1,i_2$ and also $c_{11}=0$. In the case $a_1 + a_2 \ge 1$,
%$tr(\sigma)$ is minimized by $4d_{00}=4(c_{10}+c_{01}-c_{00})=a_1+a_2-1$.
%Then $c_{11} = - d_{00}$ and  $d_{10}=d_{01}=d_{11}=-d_{00}/3$.
The upper bound on the Global Robustness of Entanglement is
thus given by:
\begin{align}
E_{min} = max\{0,|a_1|+|a_2|-1\}.
\end{align}

\subsubsection{2 Qubits: Lower Bound}
Here we will derive a lower bound on the Global Robustness of
Entanglement according to Eq. \ref{eq:dualopt}. We will see
that this lower bound coincides with the upper bound derived in the previous
section. First, one may restrict the matrices $\eta_{\alpha}$ to have the same symmetries as $\rho$:
\begin{equation}
\eta_{\alpha} = \sum_{i_1,i_2 = 0}^1 c_{i_1 i_2}^{(\alpha)} K_1^{i_1}K_2^{i_2}.
\end{equation} 
Since the partial transposes $\Gamma_1$ and $\Gamma_2$ have the same
impact on the $\eta_{\alpha}$ (both change the sign of the coefficient $c_{11}$), we may simplify the problem by setting $\eta_1
= \eta_2 = \eta/2$. Again we consider only the case of positive $a_i$'s. The dual problem can now be formulated as the
following eigenvalue problem in the form of a linear program:
% \begin{align}
% maximize~ & \mu_0 + \mu_1 a_1 + \mu_2 a_2 \\
% \label{eq:cs2LP1}
% subject~to~& \mu_0 \mathbb{1} + \mu_1 K_1 + \mu_2 K_2 + \eta^{\Gamma_1} \le 0,~ \eta \ge 0,~ \eta^{\Gamma_1} \le \mathbb{1}
% \end{align}
\begin{align}
max[\mu_0 + \mu_1 a_1 + \mu_2 a_2: \mu_0 \mathds{1} + \mu_1 K_1 
+ \mu_2 K_2 + \eta^{\Gamma_1} \le 0,~ \eta \ge 0,~ \eta^{\Gamma_1} 
\le \mathds{1}].
\end{align}

Besides the trivial solution (all variables equal zero), a little thought shows that the above system of inequalities is fulfilled by $-\mu_0 = \mu_1 = \mu_2 = 1$ and $c_{00}=-c_{10}=-c_{01}=c_{11}=1/2$. Thus:
\begin{align}
E_{min} = max\{0,|a_1| + |a_2| -1\}.
\end{align} 
which coincides with the upper bound presented in the previous
subsection.
\subsection{3 Qubits}
If the goal of the experiment was the creation of a triangle cluster state, the observables are naturally $K_1 = X\otimes Z\otimes Z$, $K_2=Z\otimes X\otimes Z$, and $K_3=Z\otimes Z\otimes X$ with measurement outcomes $a_i = tr(K_i \rho)$, $i \in \{1,2,3\}$. Then, one finds a solution similar to the 2-qubit case, in the sense that it only depends on the two largest measurement outcomes: 
\begin{align}
        E_{min} = max\{0,|a_1|+|a_2|+|a_3|-min(|a_1|,|a_2|,|a_3|)-1\}.
\end{align}
\subsection{4 Qubits}
Let us now consider the case, where the goal of the experiment 
was the creation of a 4-qubit cluster state associated with a 
square lattice graph (box cluster state). Then the four generators 
of the corresponding stabilizer group are given by $K_1 = X \otimes 
Z \otimes \mathds{1} \otimes Z$, $K_2 = Z \otimes X \otimes Z 
\otimes \mathds{1}$, $K_3 = \mathds{1} \otimes Z \otimes X 
\otimes Z$, $K_4 = Z \otimes \mathds{1} \otimes Z \otimes X$.
%\begin{align}
%K_1 = X \otimes Z \otimes \mathbb{1} \otimes Z
%\\
%K_2 = Z \otimes X \otimes Z \otimes \mathbb{1}
%\\
%K_3 = \mathbb{1} \otimes Z \otimes X \otimes Z
%\\
%K_4 = Z \otimes \mathbb{1} \otimes Z \otimes X
%\end{align} 
The measurement outcomes are denoted by $a_i = tr(\rho K_i)$ 
for $i \in \{1,\dots,4\}$ and are assumed to be non-negative.
Thus, the problem for the square lattice case reads:
% \begin{align}
% maximize~& E_{min} = \mu_0 + \mu_1 a_1 + \mu_2 a_2 + \mu_3 a_3 + \mu_4 a_4
% \\
% subject~to~&
% %\label{eq:cs4cond1}
% \sum \eta_{\alpha}^{\Gamma_\alpha} + \sum_{i=0}^{3} \mu_i K_i \le 0,~
% %\label{eq:cs4cond2}
% -\eta_{\alpha} \le 0 ,~
% %\label{eq:cs4cond3}
% \sum \eta_{\alpha}^{\Gamma_\alpha} - 1 \le 0  
% \end{align}

\begin{align}
max[\mu_0 + \sum_{i=1}^{4} \mu_i a_i:
\sum \eta_{\alpha}^{\Gamma_\alpha} + \sum_{i=0}^{3} \mu_i K_i \le 0,~
\eta_{\alpha} \ge 0 ,~
\sum \eta_{\alpha}^{\Gamma_\alpha} \le 1  ]
\end{align}

where $K_0 = \mathds{1}$. The matrices $\eta_\alpha$ are restricted to:
%\begin{align}
%\label{eq:cs4eta}
$\eta_\alpha = \sum_{i_1, \dots, i_4=0}^1 c_{i_1 \dots i_4}^{(\alpha)} K_1^{i_1}
\dots K_4^{i_4}$.
%\end{align}
The partial transposes are therefore given by:
$\eta_\alpha^{\Gamma_\alpha} = \sum_{i_1, \dots, i_4=0}^1 (-1)^{i_1 \sum_{N_\alpha} i_{N_\alpha}}
c_{i_1 \dots i_4}^{(\alpha)} K_1^{i_1}
\dots K_4^{i_4}$.

Even though, this translates to a system of inequalities which looks rather complex, one may realize
easily that the solution of the two-qubit case represents also a solution
for this system. This means, $\mu_0=-1$, $\mu_1 = \mu_2 = 1$ and $\mu_3 = \mu_4 = 0$, and regarding the $\eta_{\alpha}$ one obtains $c_{0000}^{(1)} = c^{(1)}_{1100}
= \frac{1}{4}$, $c_{1000}^{(1)}=c_{0100}^{(1)} = -\frac{1}{4}$
% \begin{align}
% &c_{0000}^{(1)} & = & \frac{1}{4}
% \\
% &c_{1000}^{(1)} & = & -\frac{1}{4}
% \\
% &c_{0100}^{(1)} & = &-\frac{1}{4}
% \\
% &c_{1100}^{(1)} & = & \frac{1}{4}
% \\
% &c_{j_1 j_2 j_3 j_4}^{(1)} & = & 0~\text{else}
% \end{align}
and $\eta_2 = \eta_1$, $\eta_3=\eta_4=0$. One may easily check, that another solution is given by the following set of parameters: $\mu_0=-5,~\mu_1 = 2,~\mu_2 =2 ,~\mu_3 = 2,~\mu_4 = 2$.
%\be
%\mu_0 & = & -5
%\\
%\mu_1 & = & 2
%\\
%\mu_2 & = & 2
%\\ 
%\mu_3 & = & 2
%\\
%\mu_4 & = & 2
%\ee
and the coefficients of the operator sum representation of the $\eta_{\alpha}$
are listed in Tab. \ref{tab:eta4}.
%\begin{table}
%\caption{Coefficients for the operator sum representation of the operators
%$\eta_{\alpha}$}
%\label{tab:eta4}
%\begin{tabular}{|c|c|c|c|c|}
%\hline
%$\alpha$ & 1 & 2 & 3 & 4
%\\
%\hline
%$c_{0000}^{(\alpha)}$ &  3/16 &  3/16 &  3/16 &  3/16
%\\
%$c_{1000}^{(\alpha)}$ & -1/16 & -2/16 &  1/16 & -2/16
%\\
%$c_{0100}^{(\alpha)}$ & -2/16 & -1/16 & -2/16 & 1/16
%\\
%$c_{0010}^{(\alpha)}$ &  1/16 & -2/16 & -1/16 & -2/16
%\\
%$c_{0001}^{(\alpha)}$ & -2/16 &  1/16 & -2/16 & -1/16
%\\
%$c_{1100}^{(\alpha)}$ &  1/16 & 1/16  & -1/16 & -1/16
%\\
%$c_{1010}^{(\alpha)}$ &  -3/16 & 1/16 & -3/16 & 1/16
%\\
%$c_{1001}^{(\alpha)}$ &  1/16 & -1/16 & -1/16 & 1/16
%\\
%$c_{0110}^{(\alpha)}$ & -1/16 & 1/16 & 1/16 &  -1/16
%\\
%$c_{0101}^{(\alpha)}$ & 1/16 & -3/16 & 1/16 & -3/16
%\\
%$c_{0011}^{(\alpha)}$ & -1/16 & -1/16 & 1/16 & 1/16
%\\
%$c_{1110}^{(\alpha)}$ & 2/16 & -1/16 & 2/16 & 1/16
%\\
%$c_{1101}^{(\alpha)}$ & -1/16 &  2/16 & 1/16 & 2/16
%\\
%$c_{1011}^{(\alpha)}$ &  2/16 &  1/16 & 2/16 & -1/16
%\\
%$c_{0111}^{(\alpha)}$ & 1/16  &  2/16 & -1/16 & 2/16
%\\
%$c_{1111}^{(\alpha)}$ &  -1/16 &  -1/16&  -1/16&  -1/16
%\\
%\hline
%\end{tabular}
%\end{table}
Summarizing these results gives:
\be
E_{min} = max[0,~(|a_1|+|a_2|-1),~2(|a_1|+|a_2|+|a_3|+|a_4|)-5]
\ee
\section{Quality of the estimate and local statistics}
\label{sec:Quality}
In order to check the usefulness of the obtained bounds, we 
compare these bounds with exact values for simulated noisy 
cluster states. We assume that after a perfect cluser state 
was created, the qubits are subject to local dephasing for a 
certain time (here we assume $\unit[10]{ms}$). Then, the system 
is described by the following master equation:
\begin{align}
\dot{\rho} = \frac{\gamma}{2}\sum_i (Z_i \rho Z_i - \rho)
\end{align}
 where $\gamma$ is the dephasing-rate, which we take to be $(\unit[10]{s})^{-1}$.
%\begin{figure}
%\includegraphics{comparison.pdf}
%\end{figure}
A comparison between exact values of the Global Robustness with 
our estimate is given in Tab. \ref{tab:comparison}. It shows that 
the estimate deviates only a few per cent from the exact value.
\begin{table}
\begin{tabular}{|c|c|c|c|}
\hline
No. Qubits & exact value & estimated value & relative deviation
\\
\hline
2 & 0.8142 & 0.8097 & 0.0055
\\
3 & 0.8185 & 0.8097 & 0.0108
\\
4 & 2.2995 & 2.2387 & 0.0264
\\
\hline
\end{tabular}
\caption{Comparison of the Global Robustness of Entanglement (GRE) for 2, 3, and 4 qubit noisy cluster states with an estimate of the GRE from measurements of the generators of the stabilizer group only}
\label{tab:comparison}
\end{table}
It is obvious that the bounds can be improved by considering any additional information on the density matrix. When one performs measurements on distant parties, the observables such as $\langle X \otimes X\rangle$ must be gained from local measurements. The entanglement depends mainly on the correlations, as we have seen in the above comparison. However, the obtained local statistics may be used to improve the bounds. Consider the measurements $\langle X \otimes X\rangle = 0.9$, $\langle Z \otimes Z\rangle = 0.7$, $\langle Z \otimes \mathds{1} \rangle=0$, $\langle \mathds{1} \otimes Z \rangle=0.25$. In this example, the GRE yields 0.6 when one considers only the XX- and ZZ-observables. Taking into account the local statistics improves the GRE by more than ten per cent to 0.6671. This example shows that one should use all available information the measurements provide to obtain optimal bounds on entanglement.

\section{Conclusion and outlook}
In conclusion, we investigated how to obtain lower bounds on 
the fidelity and robustness measures from partial information 
on the density matrix of a multi-partite system. We utilized 
the symmetries of the stabilizer group to formulate the problem
as a linear program, which can be treated analytically as well 
as numerically. Analytical solutions were obtained for two, 
three, and four qubit-systems. This method is of particular 
interest for experiments, since the number of measurement 
settings grows only linearly in the number of qubits, whereas
full-state tomography requires an exponential number of settings. A comparison of the obtained bounds with exact values of the Global Robustness shows that the difference is in the order of only a few per cent. 

For the future, it will be interesting to investigate if analytical solutions to our approach can be found for systems with an arbitrary number of constituents.

\section*{Acknowledgement}
HW thanks Fernando Brand\~{a}o, Animesh Datta, Masaki Owari, Alex Retzker 
and Shashank Virmani for very helpful discussions. This work was 
supported by the Deutsche Forschungsgemeinschaft, the EU Integrated 
Project "Qubit Applications", the EPSRC QIP-IRC and the EU STREP
project HIP.

\bibliographystyle{tMOP}
%\bibliography{pqe09_ref}

\appendix

\section{Solution to the dual problem for a 4 qubit measurement}

\begin{table}[h]
\begin{tabular}{|c|c|c|c|c||c|c|c|c|c|}
\hline
$\alpha$ & 1 & 2 & 3 & 4 & $\alpha$ & 1 & 2 & 3 & 4
\\
\hline
$c_{0000}^{(\alpha)}$ &  3/16 &  3/16 &  3/16 &  3/16 & $c_{0110}^{(\alpha)}$ & -1/16 & 1/16 & 1/16 &  -1/16 \\
$c_{1000}^{(\alpha)}$ & -1/16 & -2/16 &  1/16 & -2/16 & $c_{0101}^{(\alpha)}$ & 1/16 & -3/16 & 1/16 & -3/16 \\
$c_{0100}^{(\alpha)}$ & -2/16 & -1/16 & -2/16 & 1/16 & $c_{0011}^{(\alpha)}$ & -1/16 & -1/16 & 1/16 & 1/16 \\
$c_{0010}^{(\alpha)}$ &  1/16 & -2/16 & -1/16 & -2/16 & $c_{1110}^{(\alpha)}$ & 2/16 & -1/16 & 2/16 & 1/16 \\
$c_{0001}^{(\alpha)}$ & -2/16 &  1/16 & -2/16 & -1/16 & $c_{1101}^{(\alpha)}$ & -1/16 &  2/16 & 1/16 & 2/16 \\
$c_{1100}^{(\alpha)}$ &  1/16 & 1/16  & -1/16 & -1/16 & $c_{1011}^{(\alpha)}$ &  2/16 &  1/16 & 2/16 & -1/16 \\
$c_{1010}^{(\alpha)}$ &  -3/16 & 1/16 & -3/16 & 1/16 & $c_{0111}^{(\alpha)}$ & 1/16  &  2/16 & -1/16 & 2/16 \\
$c_{1001}^{(\alpha)}$ &  1/16 & -1/16 & -1/16 & 1/16 & $c_{1111}^{(\alpha)}$ &  -1/16 &  -1/16&  -1/16&  -1/16
\\
\hline
\end{tabular}
\caption{Coefficients for the operator sum representation of the operators
$\eta_{\alpha}$, where the goal of the experiment is a box cluster state}
\label{tab:eta4}
\end{table}

\section{Proof for the fidelity of noisy cluster states}
\label{sec:proof-for-fidelity}
Let $a_i = tr(K_i \rho)$, $i \in \{1,\dots,N\}$ be the
mean values of the stabilizer operators. Twirling over the stabilizer group
allows us to restrict to states $\rho = \frac{1}{2^N} \sum_{i_1,\cdots,i_N=0}^1 c_{i_1 \dots i_N} K_{1}^{i_1} \cdots K_{N}^{i_N}$.
%\be
%\rho = \frac{1}{2^N} \sum_{i_1,\cdots,i_N=0}^1 c_{i_1 \dots i_N} K_{1}^{i_1} \cdots K_{N}^{i_N}
%\ee
with eigenvalues $
%\be
\lambda_{j_1, \dots, j_N} (\rho) = \frac{1}{2^N} \sum_{i_1,\cdots,i_N=0}^1 (-1)^{\sum_m i_m j_m} c_{i_1 \dots i_N}$.
%\ee
Furthermore, the target state may be written as:
%\be
$|\phi\rangle\langle\phi| = \frac{1}{2^N} \sum_{i_1,\cdots,i_N=0}^1 K_{1}^{i_1} \cdots K_{N}^{i_N}$.
%\ee
\\
\textit{Primal problem:} Now we choose the coefficients $c_{i_1 ... i_N} = \sum_{k=1}^N
i_k a_k -\sum i_k + 1$.
%\be
%c_{i_1 ... i_N} = \sum_{k=1}^N
%i_k a_k -\sum i_k + 1
%\ee
One may assume, that the measurement outcomes $a_i$ are positiv. Any other choice could be achieved by local rotations of the qubits. Then, $\rho \ge 0$ follows from the fact that $\rho$ is diagonal in the stabilizer basis, and all coefficients are positive.
Because $\forall~m:$ $K_m^2=1$, we find
\be
tr(|\phi\rangle\langle\phi| \rho) & = & \frac{1}{2^N} \sum_{i_1,\dots,i_N=0}^1 c_{i_1 ... i_N}
\\
& = &  \frac{1}{2^N} \sum_{i_1,\dots,i_N=0}^1 (\sum_{k=1}^N
i_k a_k -\sum i_k + 1)
= \frac{1}{2}(\sum_k a_k - N + 2)
\ee
\textit{Dual problem:} The dual problem may be solved by the choice $\lambda_0
= -N/2+1$ and $\lambda_i = 1/2$ for $i \ge 1$. In order to check the validity
of this solution, one must prove that $\chi - \Xi
 \ge 0$ with $\chi = |\phi\rangle
\langle\phi|$ and $\Xi= \frac{1}{2}\sum_{i=1}^N K_i+(1-\frac{N}{2}) \mathds{1}$.
The eigenvalues of $\Xi$ are given by
\be
\lambda_{j_1 \dots j_N} (\Xi) = -\frac{N}{2}+1+\frac{1}{2}\sum_{i=1}^N (-1)^{j_i}
\ee
This means $\lambda_{0 \dots 0} (\Xi) = 1$. Since $\lambda_{0 \dots 0}(\chi) = 1$, we have $\lambda_{0 \dots 0}(\chi-\Xi) = 0$. It is easy to see that
$\lambda_{j_1 \dots j_N} (\Xi) \le 0$ for $(j_1, \dots, j_N) \ne (0,\dots,0)$,
and $\lambda_{j_1 \dots j_N} (\chi) = 0$ for these indices, thus $\chi-\Xi \ge 0$.

\end{document}